\documentclass[pre,twocolumn,superscriptaddress,showpacs,floatfix]{revtex4}
\usepackage{graphicx}
\usepackage{bm}
\usepackage{dcolumn}
\usepackage{amsmath}
\usepackage{amssymb}
\usepackage{subfigure}

\begin{document}

\title{Control of wavepacket spreading in nonlinear finite disordered lattices}

\author{Rodrigo A. Vicencio}

\affiliation{Departamento de F\'isica, Facultad de Ciencias, Universidad de Chile, Santiago, Chile}

\author{Sergej Flach}

\affiliation{Max-Planck-Institut f\"ur Physik komplexer Systeme, N\"othnitzer Str. 38, D-01187 Dresden, Germany}

\date{\today}

\begin{abstract}
In the absence of nonlinearity all normal modes (NMs) of a chain with disorder are spatially localized (Anderson localization). 
We study the action of nonlinearity, whose strength is ramped linearly in time.
It leads to a spreading of a wavepacket due to interaction with and population of distant NMs. Eventually the nonlinearity induced 
frequency shifts take over, and the wavepacket becomes selftrapped. On finite chains a critical ramping speed is obtained, which 
separates delocalized final states from localized ones.
The critical value depends on the strength of disorder and is largest when the localization length matches the system size.  

\end{abstract}

 \maketitle

\section{Introduction}

Spatial discreteness and nonlinearity have been probed recently within the context of Bose-Einstein condensates (BEC) in optical 
lattices~\cite{morsch06rmp} and propagation of light in nonlinear optical waveguide arrays~\cite{pr}, to name a few. The balance 
between these ingredients allows for the excitation of localized structures known as \textit{discrete breathers/solitons}~\cite{db}. 
Localization is due to nonlinearity induced frequency shifts, which tune a localized excitation out of resonance with the surrounding 
nonexcited extended lattice modes, and is also known as \textit{selftrapping}.

The normal modes (NMs) of a one-dimensional \textit{linear} chain with uncorrelated random potential are spatially localized 
(Anderson localization). Therefore any wavepacket, which is initially localized,
remains localized for all time \cite{anderson}. Anderson localization is harvesting on destructive interference and phase coherence, 
despite the fact, that the frequency of a localized NM is not tuned out of resonance with other NMs. From the experimental point of 
view, disorder can be easily implemented in BEC's by tuning the wavelength of laser beams~\cite{difbeams} or by superimposing a 
speckle pattern~\cite{speckle}. Recently experiments have been reported, which observe Anderson localization for noninteracting 
BECs \cite{becexp}. In optics, disorder can be implemented in waveguide arrays by altering the fabrication process \cite{Led2D,1Dexp} 
or, in optically induced lattices, by adding a speckle beam \cite{naturesegev}.

Nonlinearity induces interaction between NMs, and frequency shifts. While interaction favours delocalization 
\cite{Shep93Mol98,PS08,sfdokcs08}, frequency shifts may lead to selftrapping and again to 
localization \cite{kkfa08}. Continuation of NMs into the nonlinear regime may keep localization, but also delocalize excitations 
via resonances \cite{kopiaubry1}. Experimental studies of the combined action of nonlinearity and disorder are possible both in 
BEC systems, as well as in optics \cite{Led2D,1Dexp,naturesegev,schulte}.

According to Ref.~\cite{sfdokcs08}, any initial wavepacket and a fixed value of the nonlinearity parameter define three regimes - weak, 
intermediate and strong nonlinearity. The weak nonlinearity regime is characterized by Anderson localization on potentially large 
time scales, and subsequent detrapping and spreading. The intermediate regime is yielding spreading from scratch. The spreading 
continues despite the fact that the nonlinear frequency shifts weaken, since that is balanced by the increase in the number of 
excited NMs. Finally the strong nonlinearity regime leads to partial selftrapping, i.e. a part of the wavepacket selftraps and 
does not delocalize while the remaining part spreads again. 
The spreading is universal and subdiffusive, therefore rather slow, posing challenges for experimental 
studies, where one has to compete with dissipation mechanisms which lead to dephasing.

Here we study the spreading of a wavepacket in a finite lattice by ramping the strength of nonlinearity in time. The increase of 
nonlinearity with time counteracts the above diminishing of nonlinear frequency shifts, and substantially speeds up the spreading 
process. At the same time, resonant adiabatic excitation of distant NMs can contribute to a spreading as well. That makes our 
results also easier accessible for experiments. 

\section{Model}

\subsection{Equations of motion}

We consider a discrete nonlinear Schr\"odinger (DNLS) model which describes the propagation of light in nonlinear waveguide arrays 
or the evolution of a BEC in a periodic optical potential~\cite{pr,morsch06rmp}. The Hamiltonian
\begin{equation*}
\mathcal{H}=- \sum_l \left[\epsilon_l |\psi_l|^2
+ (\psi_{l+1}\psi_{l}^{*} +\psi_{l+1}^{*}\psi_{l}) +\frac{\beta}{2} |\psi_l|^{4}\right].
\end{equation*}
The equations of motion are generated with $\partial \psi_l/ \partial z = \partial \mathcal{H}/
\partial (i \psi^{*}_l)$:
\begin{equation}
-i \frac{\partial \psi_{l}} {\partial z} =
\epsilon_l \psi_l +\left(\psi_{l+1}+\psi_{l-1}\right)+\beta(z)|\psi_{l}|^2 \psi_{l}\;,
\label{eq}
\end{equation}
where $\psi_l(z)$ is a complex wave amplitude at site $l$. $z$  corresponds to the dynamical variable 
(propagation coordinate for photons or time for atoms). The nonlinear coefficient $\beta \equiv \beta(z)\equiv c_0 z^{\mu}$ increases 
with $z$ ($\mu > 0$). We will numerically investigate linear ramping $\mu=1$ with velocity $c_0>0$. This can be implemented for the case 
of a BEC, where the interaction between condensed atoms can be described by a single parameter, the scattering length, which is 
proportional to an external applied magnetic field. This parameter can be adjusted via Feshbach resonances and therefore the 
nonlinear interaction between particles can be tuned and controlled in time~\cite{science02}. In optical systems, the nonlinearity 
can be adjusted in the propagation direction ($z$) in the fabrication process of laser-written waveguide arrays~\cite{jena} or 
by controlling the doping concentration in photovoltaic samples~\cite{dk}.

The random on-site energies $\epsilon_l$ are chosen uniformly from the interval $\left[-\frac{W}{2},\frac{W}{2}\right]$, where $W$ is 
the strength of disorder. The norm $\mathcal{N}\equiv \sum_l |\psi_l|^2$ is dynamically conserved, i.e. its value does not change 
with $z$.  In our calculations $\mathcal{N}=1$. 
We note that varying the norm is strictly equivalent to varying $\beta$. For $c_0=0$, (\ref{eq}) is reduced to the linear eigenvalue 
problem $-\lambda_{\nu} A_{\nu,l} = \epsilon_l A_{\nu,l} +(A_{\nu,l+1} + A_{\nu,l-1})$. The eigenvectors $A_{\nu,l}$ are the NMs 
and the eigenvalues $\lambda_{\nu}$ their frequencies.

We analyze normalized distributions $n_l=|\psi_{l}|^2 \geq 0$ using the participation number $P=1 / \sum_{l} n_{l}^2$, which 
measures the number of the strongest excited sites, and the second moment
$m_2= \sum_{l} (l-\bar{l})^2 n_{l}$, which measures the size of the wavepacket. Here $\bar{{l}} = \sum_{l} l n_{l}$.

\subsection{The linear case $c_0=0$}

Let us first consider an infinite system. For $W=c_0=0$, solutions of Eq.~(\ref{eq}) are extended plane waves 
$u_l(z)=u_0 \exp [i(kl-\lambda_k z)]$ with the dispersion relation $\lambda_k\equiv-2\cos k$. The frequencies $\lambda_k$ are 
confined to the interval $[-2,2]$ ($k$ is related to the input angle for optics or to the quasi-momentum for BEC). The group 
velocities of the plane waves $|v_g| \leq 2$. 

For $W\neq 0,c_0=0$, the frequency spectrum is confined to $ [-2-\frac{W}{2},2+\frac{W}{2}]$. 
The width of the spectrum $\{ \lambda_{\nu} \}$ is $\Delta=4+W$. All eigenvectors will be localized in space \cite{anderson}. 
The asymptotic spatial decay of an eigenvector is given by $A_{\nu,l} \sim {\rm e}^{-\xi l}$ where 
$\xi(\lambda_{\nu}) \leq \xi(0) \approx 100/W^2$ is the localization length~\cite{loca}. The NM participation number 
$p_{\nu} = 1/\sum_l A_{\nu,l}^4$ characterizes the spatial extend - localization volume - of the NM. It is distributed around the 
mean value $\overline{p}\approx 3.6 \xi$ with variance $\approx (1.3 \xi)^2$ \cite{MIRLIN}. The average spacing of eigenvalues 
of NMs within the range of a localization volume is therefore $\overline{\Delta \lambda} \approx \Delta / \overline{p} $.

The linear case is characterized by two frequency scales: the width of the spectrum $\Delta$, and the average spacing 
$ \overline{\Delta \lambda} < \Delta$. In addition, since we will operate with finite systems, we also have two spatial scales - 
the localization length $\xi$ and the system size $N$. 

\begin{figure}[h]
\centerline{\scalebox{0.4}{\includegraphics{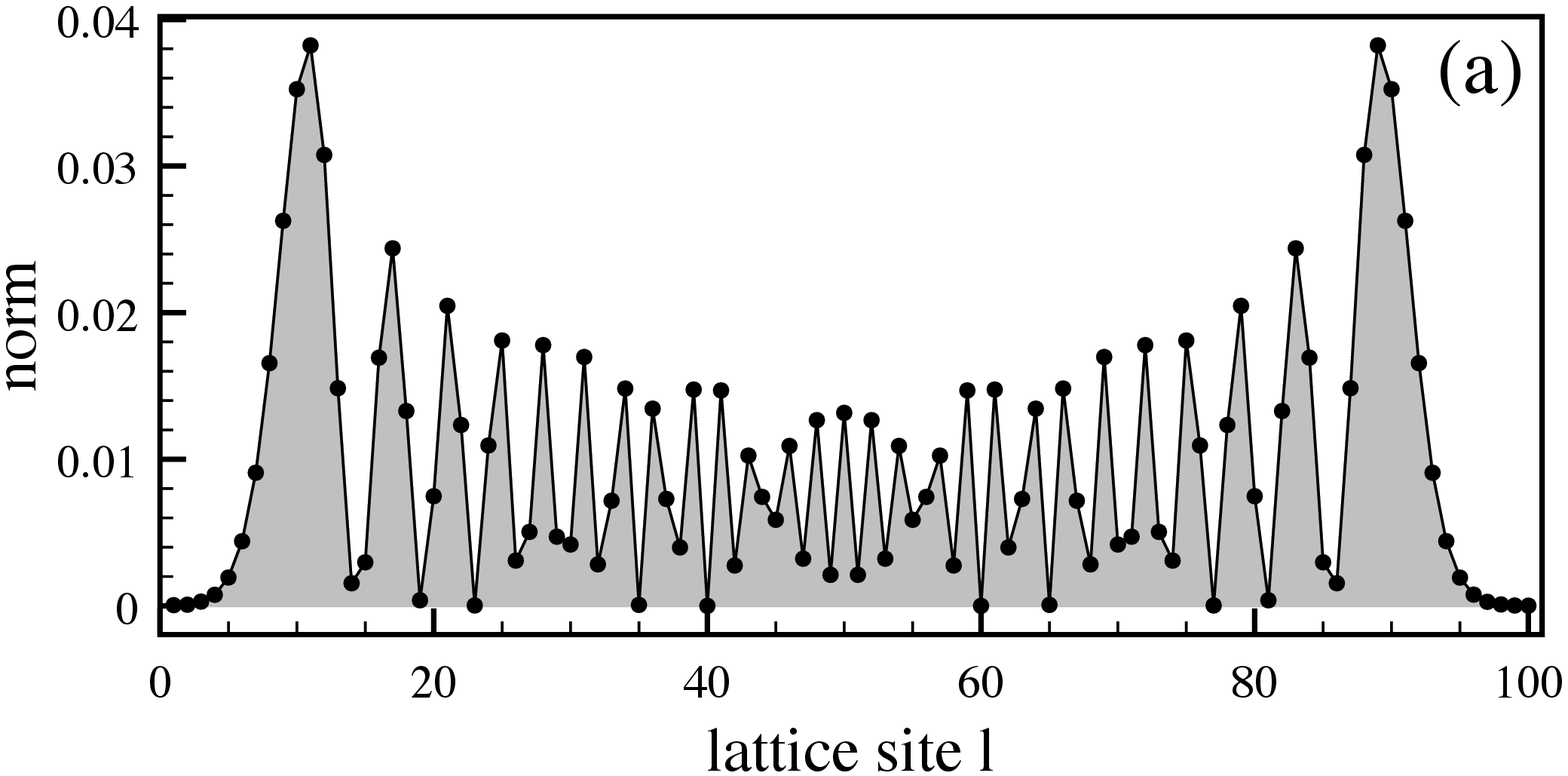}}}
\centerline{\scalebox{0.4}{\includegraphics{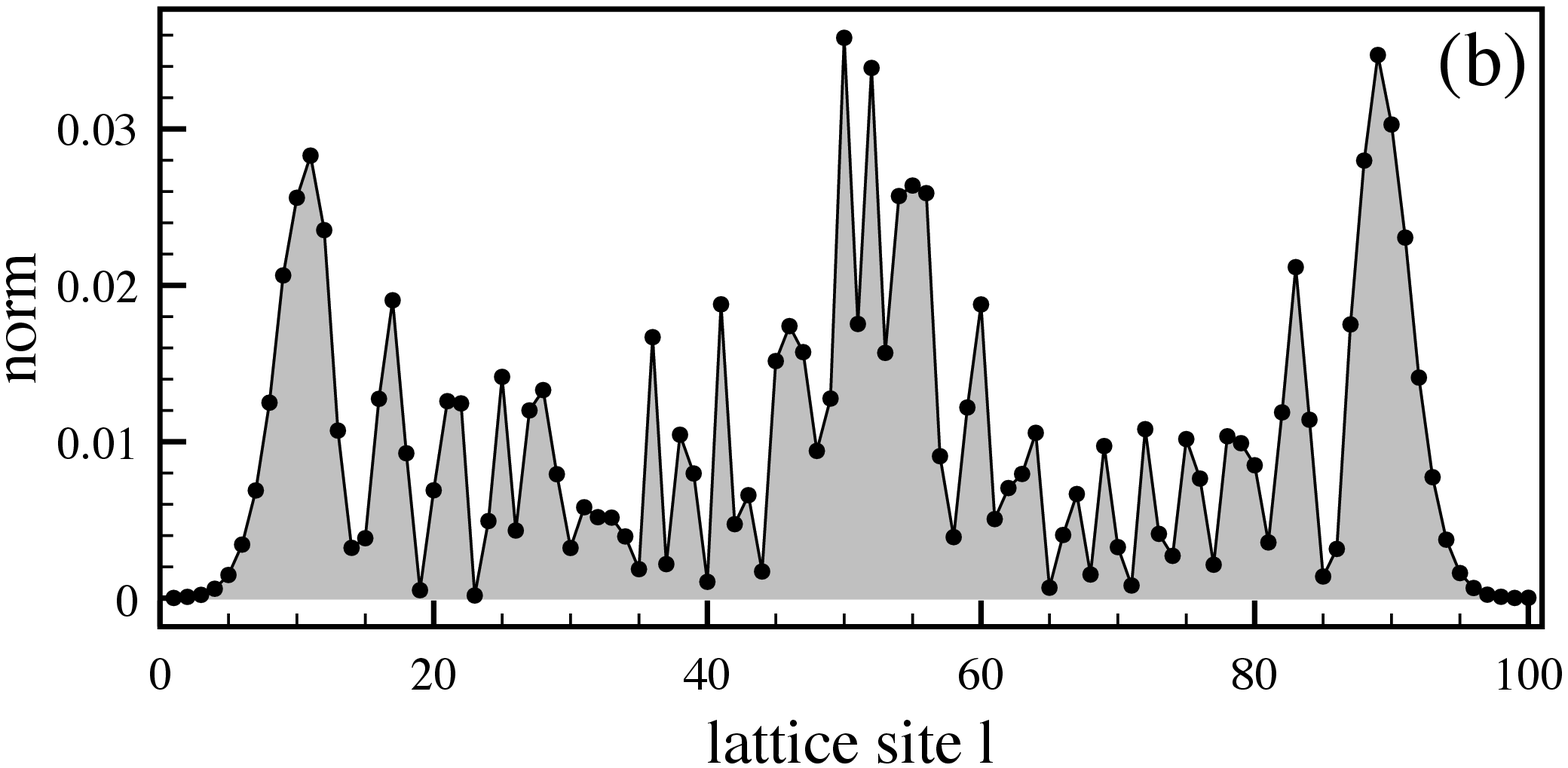}}}
\centerline{\scalebox{0.4}{\includegraphics{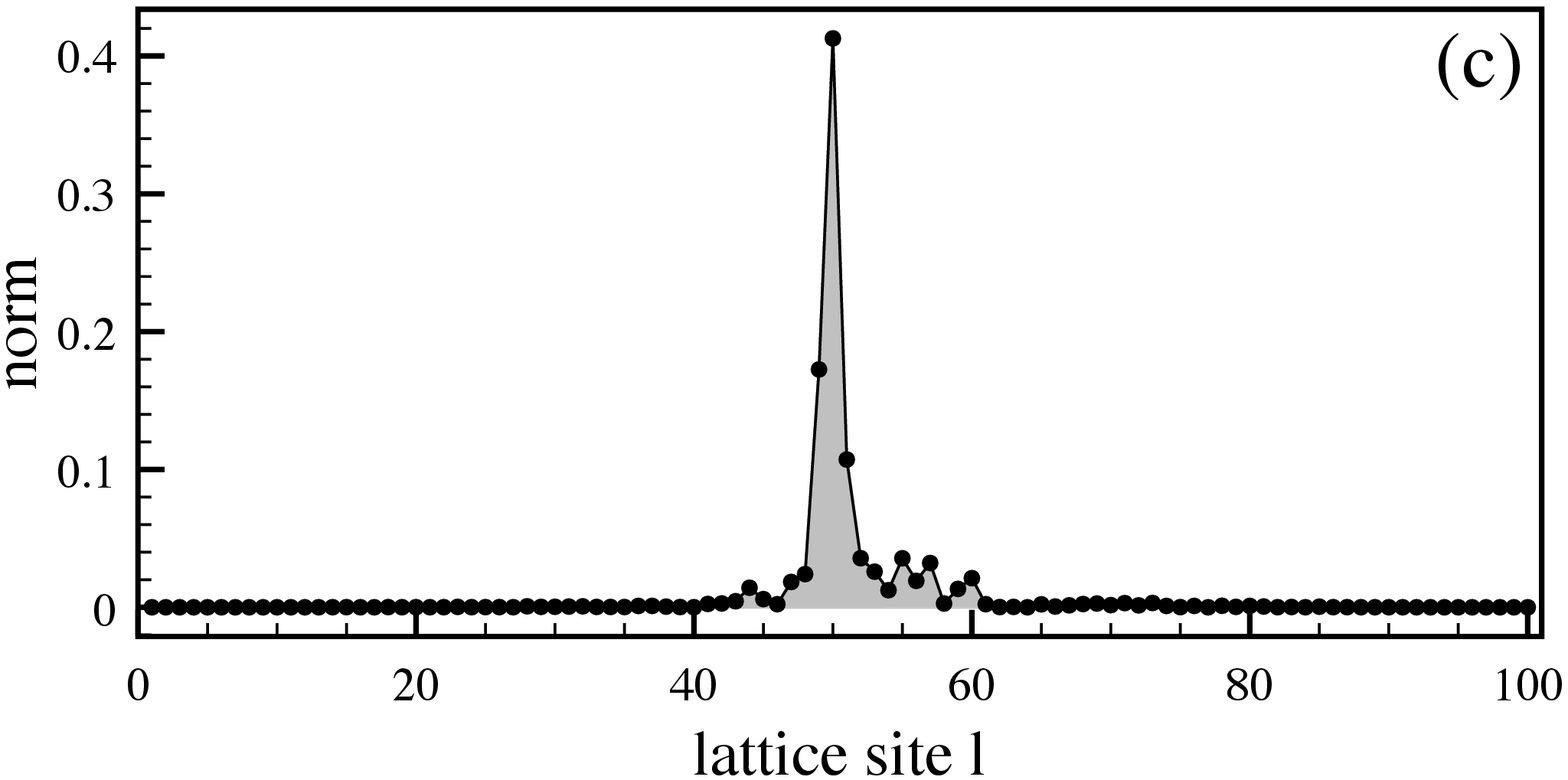}}}
\caption{{\bf Linear propagation}. Spatial profile at $z = z_{max}$ for (a) $W=0$, 
(b) $W=0.5$ and (c) $W=3$. $N=100$.}\label{linear}
\end{figure}

We launch a single site excitation in the center of our system with $N=100$, and follow its evolution until $z = z_{max}$ 
($z_{max} \equiv N/4$). For $W=0$, the fastest plane waves will reach the boundaries exactly at $z=z_{max}$. A plot of the 
corresponding norm profile is shown in Fig.\ref{linear}(a). The corresponding participation number is $P\approx 44 =0.44 N$. 
Indeed, an almost completely spread wavepacket is characterized by roughly $P=N/2$, since a strictly homogeneous distribution is 
counterbalanced by dynamical fluctuations, and therefore, on average, every second site is not excited.
%
%
Now we increase the strength of disorder $W$. For $W=1$ the localization length equals the system
size. Therefore, for $W=0.5$ [Fig.\ref{linear}(b)] we still observe almost full spreading, although the fronts propagate slower. 
For $W=3$ [Fig.\ref{linear}(c)] we observe Anderson localization - the wavepacket is confined to the localization volume which 
is of the order of 20 sites.

\begin{figure}[h]
\scalebox{0.4}{\includegraphics{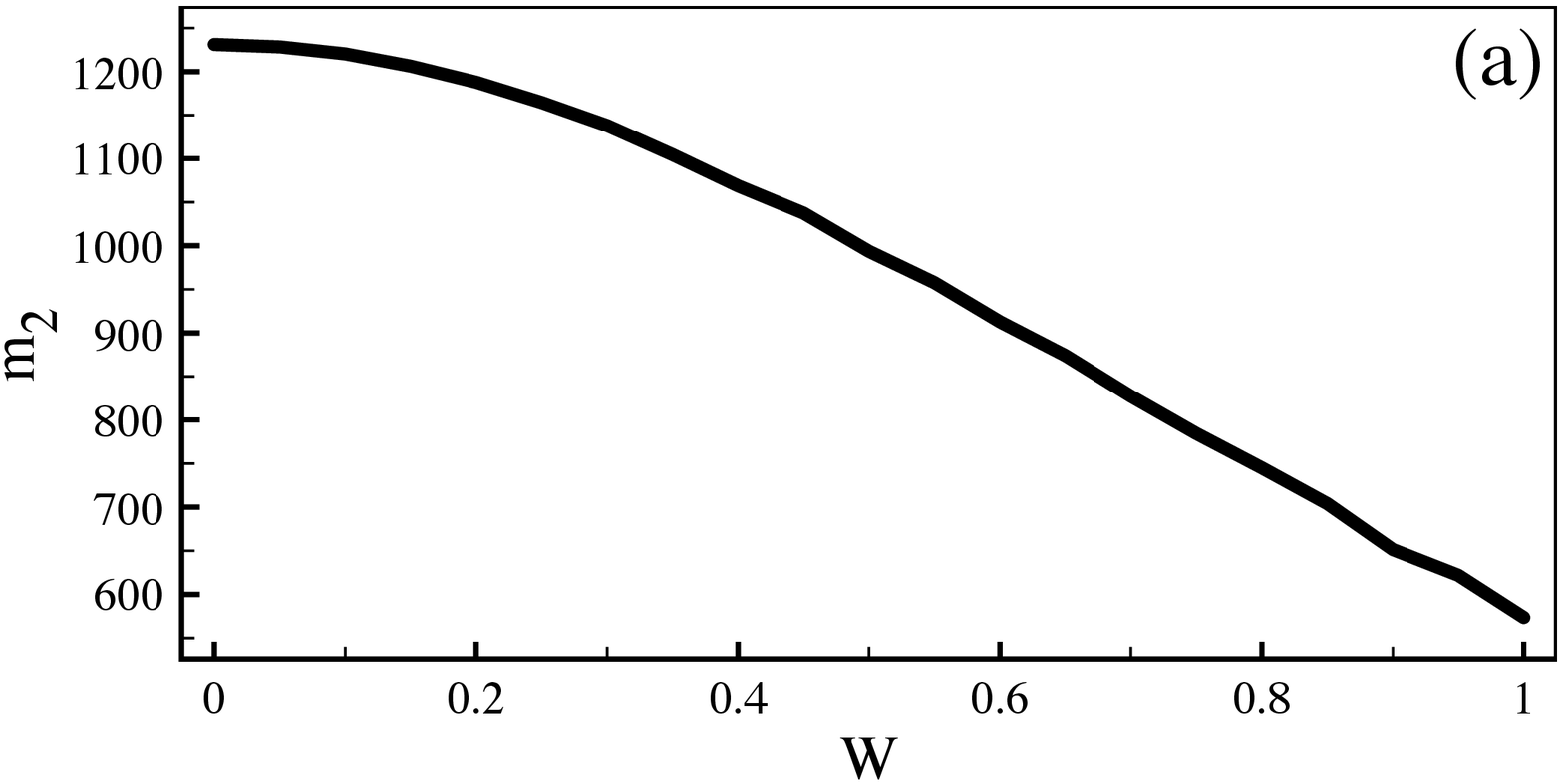}}
\scalebox{0.4}{\includegraphics{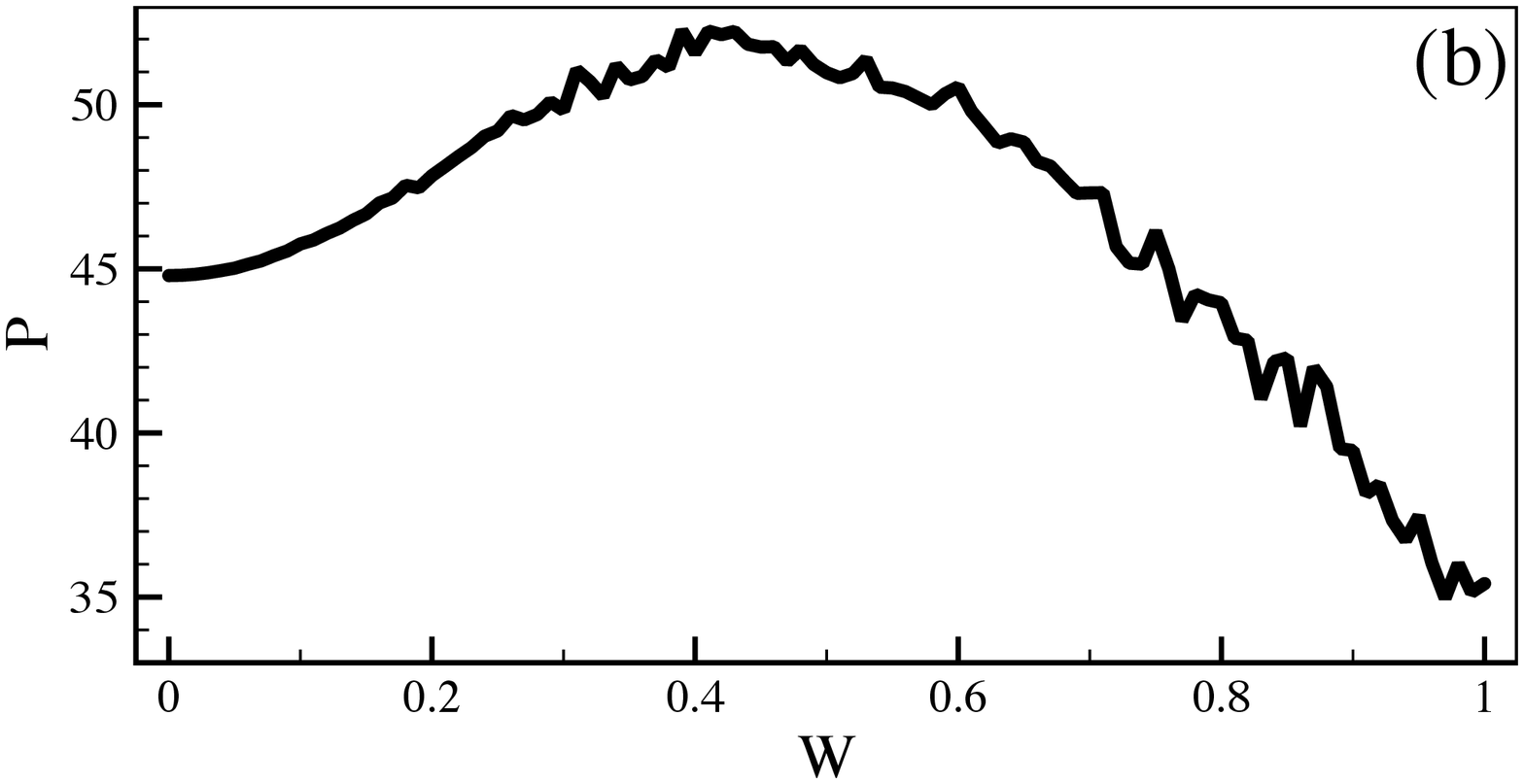}}
\caption{{\bf Linear propagation}. (a) Average second moment $m_2$ measured at time $z_{max}$ as 
a function of $W$. (b) Same as (a), but for the participation number $P$. $N=100$ and averaging was performed with 100 realizations.}
\label{linear_pm2}
\end{figure}

We performed runs for 100 realizations, and compute the average of the second moment $m_2$  and of the participation number 
$P$ at $z_{max}$ (see Fig.\ref{linear_pm2}). The second moment $m_2$ decreases with increasing disorder as expected, showing that 
waves propagate slower even within the localization volume ($W < 1$) due to disorder induced backscattering. However, the 
participation number $P$ shows a slight increase for weak disorder, with a subsequent (expected) decrease for stronger disorder. For 
the ordered case $W=0$ the chosen initial condition excites one half of all available NMs. That follows from the reflection 
invariance of the lattice around the center, where the initial wavepacket is placed. 
All NMs separate into two irreducible groups of even and odd ones, or symmetric and antisymmetric ones with respect to
reflections around the chain center. A single site excitation in the center excites only even NMs.
For weak disorder all NMs start to become 
excited. Therefore the part of the volume which is occupied by the wavepacket at some later time, is filled more homogeneously. 
Note that the effect is weak - the participation number increases by $\sim15\%$. That may be due to the fact, that at the same 
time the wavepacket spreads less effectively, as it follows from the results on the second moment.
%
%

\subsection{The nonlinear case}

The equations of motion in normal mode space read
\begin{equation}
-i \dot{\phi}_{\nu} = \lambda_{\nu} \phi_{\nu} + \beta \sum_{\nu_1,\nu_2,\nu_3}
I_{\nu,\nu_1,\nu_2,\nu_3} \phi_{\nu_1} \phi_{\nu_2}^* \phi_{\nu_3}\;
\label{NMeq}
\end{equation}
with the overlap integral
\begin{equation}
I_{\nu,\nu_1,\nu_2,\nu_3} = \sum_l A_{\nu,l} A_{\nu_1,l} A_{\nu_2,l} A_{\nu_3,l}\;.
\label{OVERLAP}
\end{equation}
The variables $\phi_{\nu}$ determine the complex time-dependent amplitudes of the NMs.  

Nonlinearity therefore induces interaction between NMs. In particular the nonlinearity renormalizes frequencies. If frequencies are 
shifted out of the spectrum of the linear equation, selftrapping occurs,
and excitations stay localized for long, may be infinite, times. We are not aware of a straightforward and unique way to calculate 
such a frequency shift for a given distribution $n_l$. Therefore we will look for suitable estimates. One possibility is to neglect 
the coupling along the lattice, and treat lattice sites as independent. Then the nonlinear frequency shift at site $l$ is 
$\delta \lambda \approx -\beta n_l$. Another possibility is to derive an effective frequency $\lambda_{eff}$ for a given state. 
We treat a state as a stationary one $\psi_{l}(z)=A_{l}\exp{(-i \lambda_{eff} z)}$. 
We insert this ansatz in (\ref{eq}), multiply this equation by $\psi_{l}^*$ and sum over all lattice sites:
\begin{equation}
\lambda_{eff}=-\sum_l \left[\epsilon_l |\psi_l|^2+\psi_{l+1}\psi_{l}^*+\psi_{l+1}^*\psi_{l}+
\beta|\psi_l|^4\right].
\label{L}
\end{equation}
At any time $z$ the effective frequency $\lambda_{eff}(z)=\mathcal{H}(z)-\beta(z)/[2P(z)]$. In practice both definitions from 
above are giving similar results, and we will mainly use (\ref{L}).

In all our simulations we use a single-site excitation as the initial condition: $\psi_l(0)=\delta_{l,l_c}$ with $l_c=N/2$, such 
that $\lambda(0)=-\epsilon_{l_c}$ and $P(0)=1$. To effectively harvest on resonances with the spectrum we have imposed that 
$\epsilon_{l_c}=-W/2$. With that, and for $c_0> 0$, the effective frequency will decrease starting from $\lambda_{eff}=W/2$ 
until it reaches the bottom of the spectrum $\lambda_{bot}$ where we stop our simulations. The bottom of the spectrum of an 
infinite system is located at $\lambda = -2-W/2$. However, for finite systems the bottom of the spectrum $\lambda_{bot} \geq -2-W/2$ 
depends on the given realization. From (\ref{L}) we see that a positive increment in $\beta$ will decrease the effective 
frequency of the system. Therefore, the state will be able to resonate with other NMs inside the spectrum. Outside of it, the 
solution transforms into a selftrapped localized state similar to a discrete soliton~\cite{Led2D,1Dexp,kopiaubry1}, which is a 
time-periodic and exponentially localized excitation~\cite{db}.

\section{Numerical results}

\begin{figure}[h]
\centerline{\scalebox{0.41}{\includegraphics{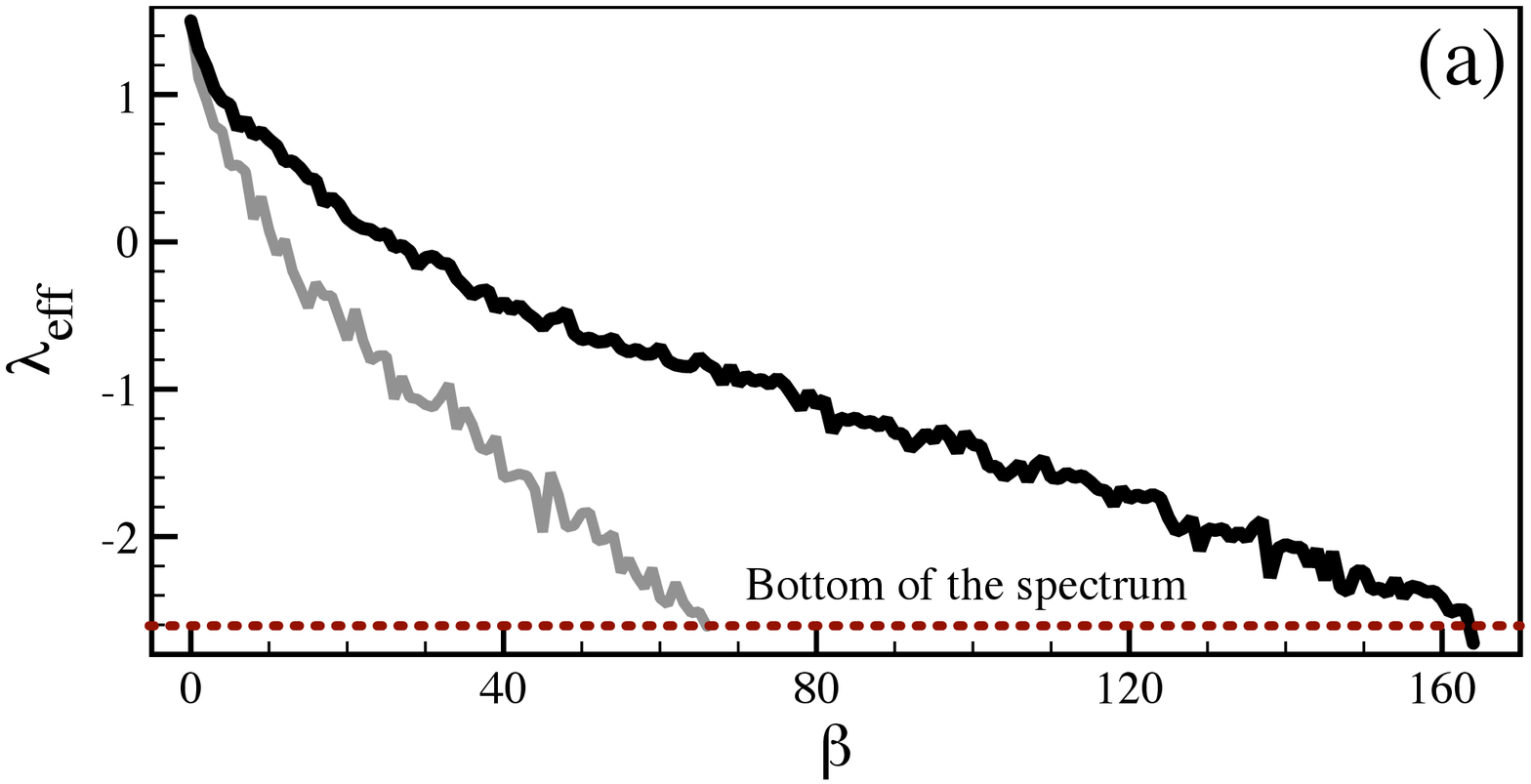}}}
\centerline{\scalebox{0.41}{\includegraphics{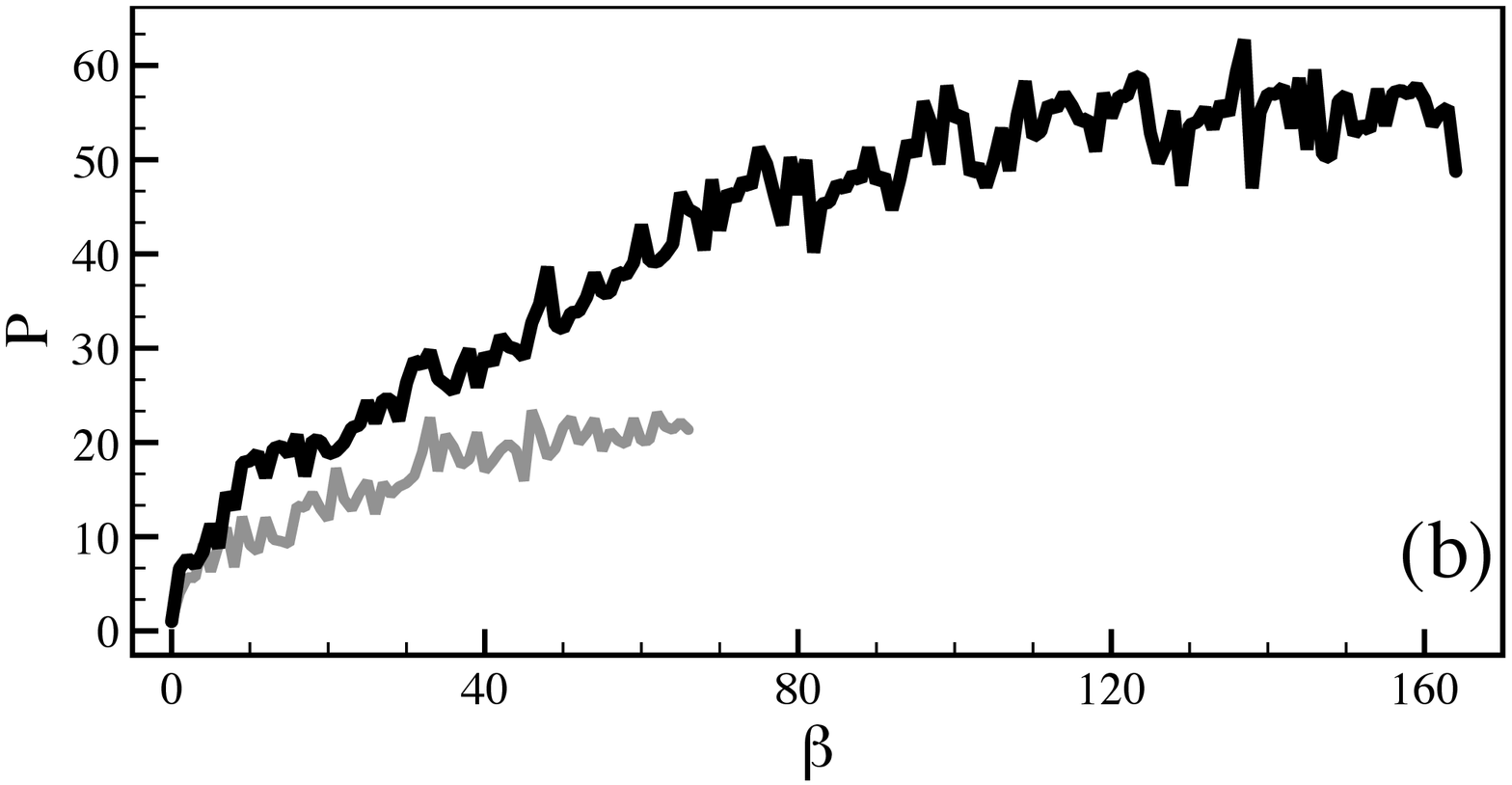}}}
\centerline{\scalebox{0.41}{\includegraphics{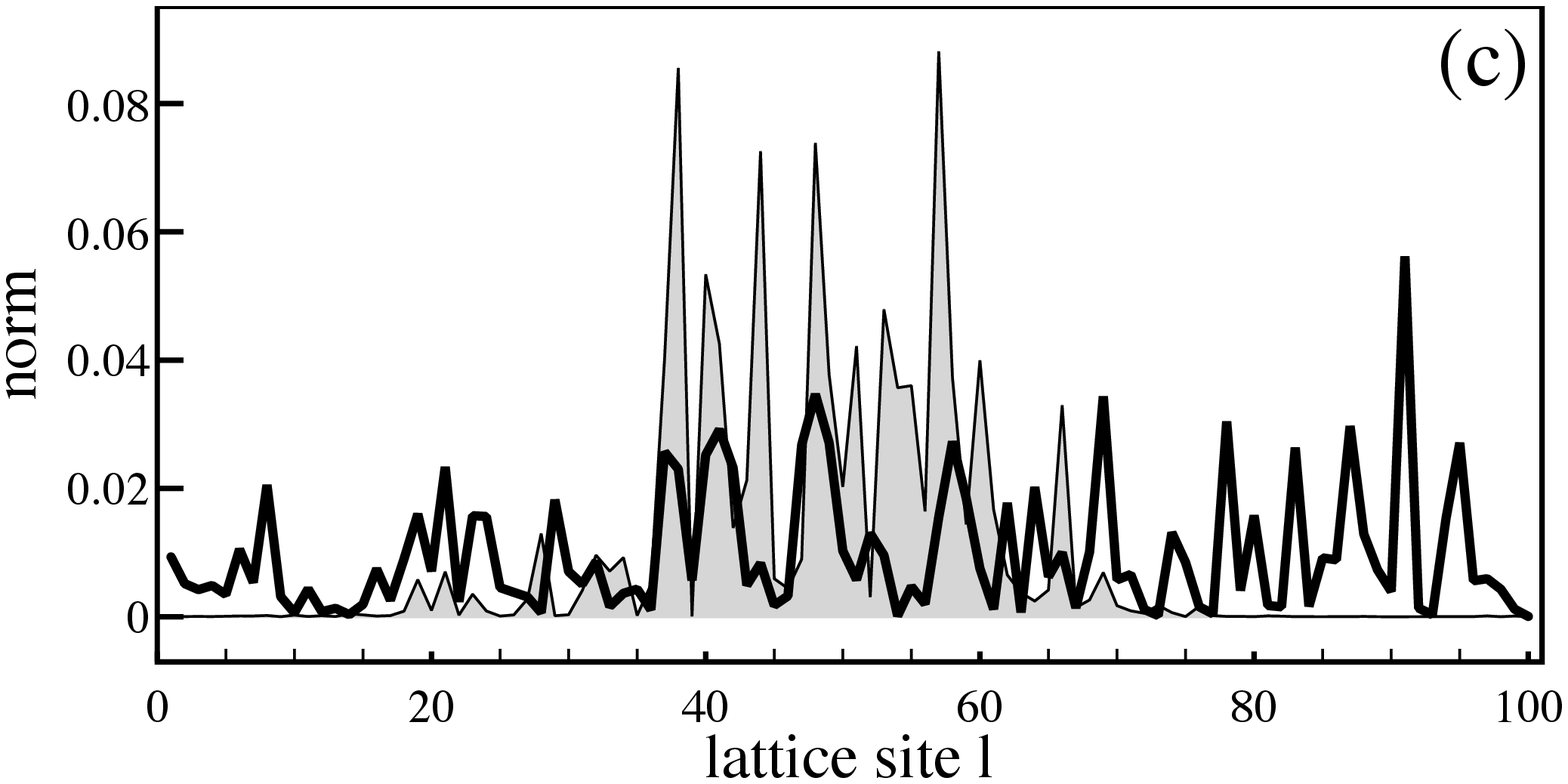}}}
\caption{{\bf Nonlinear propagation}. (a) $\lambda_{eff}$ versus $\beta$; (b) $P$ versus $\beta$; (c) final profiles. Gray and 
black curves correspond to $c_0=1$ and $0.2$, respectively. $W=3$ and $N=100$. The horizontal dashed line in (a) corresponds to 
$\lambda_{bot}$.}
  \label{ao3}
\end{figure}

In Fig.~\ref{ao3} we show the evolution of an initial single site excitation for $W=3$ and $N=100$, when $\xi < N$. For $c_0=0$, the 
excitation remains trapped due to Anderson localization, and does not reach the boundaries of the finite chain 
[see Fig.~\ref{linear_pm2}(c)]. However, for $c_0=0.2$, a slow increase of nonlinearity leads to a complete delocalization of the 
wavepacket at $z\approx 800$. Indeed, the effective frequency $\lambda$ decreases and around that time reaches the bottom of the 
spectrum. At the same time, the participation number reaches its saturation value $P\approx N/2=50$. 
For a larger ramping velocity $c_0=1$, but exactly the same disorder realization, the wavepacket does not spread over the entire 
chain. At a much shorter time $z\approx 70$ the effective frequency touches the bottom of the spectrum, the participation number 
saturates around $P\approx 20$, and the state becomes selftrapped, still occupying only 20 out of 100 sites. 
%
Further increase of the ramping velocity will make the final wavepacket more and more
localized.

\begin{figure}[h]\centerline{\scalebox{0.6}{\includegraphics{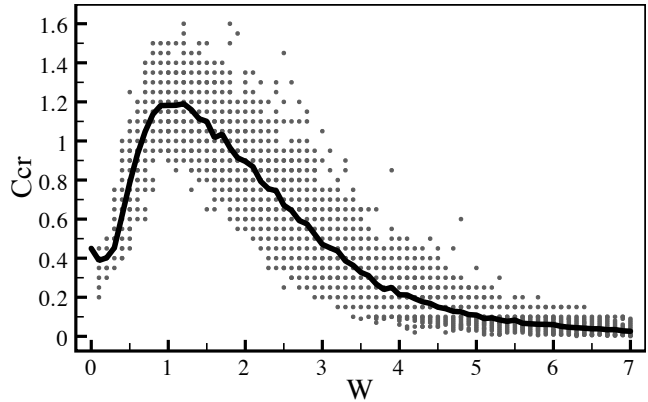}}}
\caption{Critical velocity versus degree of disorder for $100$ realizations. 
Line represents the average $\overline{c_{cr}}$. $N=100$.}
  \label{data}
\end{figure}

In order to study the dependence of the delocalization process on the disorder strength, we define our simulation scheme as follows: 
(i) we ramp the nonlinearity for a given velocity $c_0$; (ii) we stop the simulation when $\lambda_{eff}$ reaches the bottom of the spectrum 
$\lambda_{bot}$, indicating a stop of spreading due to selftrapping; (iii) we compute $P$ and if $P<40$ we decrease the velocity 
$c_0$ until $P>40$, which corresponds to a fully delocalized wavepacket. Therefore we obtain the critical velocity $c_{cr}$ for a 
given disorder realization. For $c_0 > c_{cr}$ the wavepacket does not spread over the entire chain, while for $c_0 < c_{cr}$ it 
does. We repeat the scheme for 100 different disorder realizations and obtain an average value for $\overline{c_{cr}}$. Finally we 
change the disorder strength $W$ and obtain the dependence $\overline{c_{cr}}(W)$. Results are shown in Fig.\ref{data}.
%
%

We observe two different regimes. For strong disorder $W>1$ the localization length is smaller than the system size $N=100$. The 
critical velocity $\overline{c_{cr}}$ is monotonously decreasing with increasing disorder strength. This can be expected, since 
the stronger the disorder, the more localized the NMs are, and the less is the number of other NMs a given NM can interact with 
due to nonlinearity. Therefore we need slower ramping in order to give ample time to NMs to interact.
For weak disorder $W<1$ the localization length is larger than the system size $N=100$. Therefore all NMs are essentially extended 
over the entire chain. In this regime, we observe that the critical velocity $\overline{c_{cr}}$ increases with increasing disorder 
strength. This effect may be due to the fact, that in the limit of zero disorder $W=0$, the nonlinearity induces a selective 
interaction between NMs, as mentioned above (for details of that selective interaction, see Ref.\cite{qb_dnls}). Weak but nonozero 
disorder removes the selection rules, and leads to an interaction of all modes with each other, thereby increasing the number of 
available modes (channels) by a factor of two. The nonlinearity induced spreading is more effective. In particular it can redistribute 
the energy into two times more modes.

The hallmark of the transition from weak to strong disorder is the maximum in the dependence of $\overline{c_{cr}}(W)$ at $W=1$.   
The maximum is located at $W\sim \sqrt{100/N}$. In particular, for $N=50$ it is 
located at $W\sim 1.4$ and for $N=200$ at $W\sim 0.7$. For large  $N \gg \xi$ the maximum position is therefore shifted closer to $W=0$, 
and the value in the maximum increases. The value of $\overline{c_{cr}}$ at $W=0$ can be estimated by noting that the fastest plane 
waves reach the edges of the chain at a time $z_b\approx N/4$. The time it takes to shift the frequency of the initially excited 
oscillator out of the band is $z_s \sim 4/c_0$. Equating $z_b=z_s$ we estimate $\overline{c_{cr}}(W=0) \sim 16/N$. The increase 
of this critical speed in the presence of weak disorder is due to the above discussed increase of modes (channels) available. This 
increase by a factor of two leads to a decrease of the norm in each mode also by a factor of two,  
and, consequently, implying a larger critical velocity. Therefore $z_s \sim 2/c_0$ and 
$\max (\overline{c_{cr}}(W\neq 0)) \sim 2 \overline{c_{cr}}(W=0)$ at best. This increase by a factor of two is close to the numerical 
observation (see Fig.\ref{data}).

\section{Diffusion versus resonant spreading}

Let us discuss possible mechanisms of wavepacket spreading in the regime of strong disorder, when $\xi < N$. The packet spreading beyond 
the localization volume is entirely due to the nonlinearity, which induces interactions between NMs. Assume the wavepacket has a certain 
size at some time $z$. Further spreading implies excitation of exterior NMs. Since the interaction between NMs falls off exponentially 
for distances larger than $\xi$, the relevant exterior NMs to be excited will be located in a nonexcited (cold) boundary layer outside 
the wavepacket, with a width roughly of the order of the localization volume. A given cold exterior NM can be excited coherently or 
incoherently. A coherent excitation implies a resonance with an excited NM from the wavepacket, and a corresponding resonant transfer 
of energy (as it happens during the beating of energy between two weakly coupled harmonic oscillators).
An incoherent excitation implies the absence of such resonances, and an almost random fluctuation of the NMs phases inside the 
wavepacket. It generates a random force which incoherently excites (heats) the exterior NMs. Such a spreading corresponds to a 
diffusive spreading of the wavepacket, therefore of the norm, and the energy.

\subsection{Diffusion}

The spreading of wavepackets in the presence of a constant nonzero nonlinearity strength $\beta=\beta_0$ was studied in several papers 
\cite{Shep93Mol98,PS08,kkfa08,sfdokcs08}. In particular, it was observed, that a spreading wavepacket is characterized by a growth of 
its second moment $m_2 \sim z^{\alpha}$. The exponent $\alpha = 0.33\pm 0.02 $ \cite{sfdokcs08}. A theoretical analysis showed, that 
this subdiffusive spreading is not due to resonant excitation of exterior modes, but due to incoherent heating. The origin of the 
chaotic dynamics of the wavepacket itself comes from resonant interaction of NM pairs inside the packet. The statistical analysis of 
the probabilities of such internal resonances leads to the conclusion, that the second moment obeys the following equation:
\begin{equation}
\frac{d m_2}{dt} = C(W) \beta^4 n^4\;.
\label{dif1}
\end{equation}
Here $n$ is the average norm density inside the wavepacket, and $C(W)$ an unknown function, which however, as numerical studies suggest, 
decreases with increasing $W$. Since the packet size is $\sim 1/n$, it follows that the participation number $P \sim 1/n$ and the second 
moment $m_2 \sim 1/n^2$. As a consequence, $m_2 \sim (\beta^4 z)^{1/3}$. The exponent $\alpha = 1/3$ and is in very good agreement
with the numerical studies. The nonlinearity induced frequency shift inside the packet $ \delta \lambda \sim \beta n \sim z^{-1/6}$. 
The more the packet spreads, the smaller the frequency shifts are. This weakening is counterbalanced by the increase in the size of 
the packet, so that more modes are involved, and guarantee a slow but steady subdiffusive spreading. 

Assuming, that in the present case the spreading is incoherent as well, we find with $\beta = c_0 z^{\mu}$:
\begin{equation}
m_2 \sim c_0^{4/3} z^{(4\mu+1)/3} \;,\; P \sim c_0^{2/3} z^{(4\mu+1)/6}\;.
\label{dif2}
\end{equation}
The nonlinearity induced frequency shift
\begin{equation}
\delta \lambda \sim \beta n \sim z^{(2\mu-1)/6}\;.
\label{dif3}
\end{equation}
In the present numerical studies $\mu=1$, and we find $m_2 \sim z^{5/3}$, $P \sim z^{5/6}$ and $\delta \lambda \sim z^{1/6}$.  
At variance with the case of constant nonlinearity (which yields subdiffusion with $\alpha=1/3$) we obtain superdiffusion with 
$\alpha=5/3$. This is due to the fact, that the increasing nonlinearity counterbalances the decay of the frequency shifts, and 
enhances the interaction of the NMs from the wavepacket with cold exterior NMs. The growth of the frequency shifts will finally 
lead to a selftrapped state, as observed in the numerical studies. 

Let us estimate the dependence of the participation number $P$ of the selftrapped wavepacket on $c_0$. With Eq.(\ref{dif2}) it follows
\begin{equation}
n \sim c_0^{-2/3} z^{-(4\mu+1)/6}\;.
\label{est1}
\end{equation}
Therefore the nonlinear frequency shift $\delta \lambda =\beta n \sim c_0^{1/3} z^{(2\mu-1)/6}$. When this frequency shift reaches 
some finite (disorder dependent) value at $z=z_f$, the wavepacket selftraps.
Therefore $z_f \sim c_0^{-2/(2\mu-1)}$. Finally
\begin{equation}
P(z_f) \sim \frac{1}{n(z_f)} \sim c_0^{-1/(2\mu-1)}\;.
\label{est2}
\end{equation} 
For $\mu=1$ we find that the selftrapped wavepacket has size $P \sim 1/c_0$. Increasing the ramping speed $c_0$ therefore leads 
to a smaller extend of the selftrapped wavepacket, as observed in the numerical results. We can not estimate the measured dependence 
$\overline{c_{cr}}(W)$, since we do not know the function $C(W)$ from (\ref{dif1}).

\subsection{Resonant spreading}

Let us assume that the wavepacket spreads by resonantly exciting exterior NMs. Except for exponentially small ramping velocities, 
the relevant exterior NMs will be located in a layer of the width of the localization volume. The average frequency spacing 
$\overline{\Delta \lambda}\approx \Delta / \overline{p}$ separates possible frequencies of resonant interactions, which can take 
place due to the nonlinearity induced frequency shift. Therefore the number of possible resonances is limited to the number of 
NMs within one localization volume, i.e. to the localization volume $\overline{p}$ itself. The maximum size, to which 
a wavepacket can grow, 
is then proportional to the $\overline{p}^2$. The numerical results in Fig.\ref{data} show, that for strong disorder $W =7$ the 
critical velocities are of the order of $\overline{c_{cr}} \approx 0.01$. For such a velocity the wavepacket spreads completely 
over a chain with 100 sites. The maximum localization length $\xi \approx 2$, therefore the localization volume is less than
$\overline{p}\approx 7$, and the maximum size to less than $\overline{p}^2 \approx 50$, which indicates, that resonant spreading 
is not the dominant mechanism. 

Let us assume the optimum case scenario for resonant spreading. That implies, that at a given time a resonance at the edge of the 
wavepacket can take place, and the energy is transfered much faster, than it takes to shift the frequency to the next resonance. 
Assume we start with one NM excited, with its frequency located at the edge of the spectrum. For $\mu=1$, we will hit the next 
possible resonance after time $z_1=\overline{\Delta \lambda}/c_0$. We assume, that a resonant mode is available in the localization 
volume, and is immediately excited. The norm is now equally distributed between both NMs, each of them having norm $1/2$. 
Nonlinearity is further ramped up, but in order to shift the frequency to the next resonance, a larger time $z_2=2z_1$ is needed, 
and so on. As long as the wavepacket does not selftrap, we find $z_j=jz_1$. Therefore the total time to reach the $j$-th resonance 
scales as $z \sim j^2$, the participation number $P \sim j$ grows as $P\sim \sqrt{z}$ and the second moment $m_2 \sim z$. Since 
that idealized process is already slower than the incoherent spreading due to diffusion, we expect that resonant spreading is 
weakly contributing to the numerically observed spreading.

\section{Conclusions}

We have investigated the spreading of a wavepacket in a disordered nonlinear chain, when ramping the nonlinearity in time. 
For linear ramping $\mu=1$ we find that the spreading is mostly due to incoherent diffusive processes. For an infinite chain, 
the nonlinear frequency shift always leads to a selftrapping of the wavepacket, and therefore not to a complete delocalization. 
The second moment $m_2 \sim z^{5/3}$ is predicted to grow superdiffusively, at variance to the previously studied case of 
constant nonzero nonlinearity, which yields subdiffusion. Therefore the present case is easier accessible in experiments, due 
to possible restrictions on maximum propagation times. Experiments are done with finite systems. We studied the case of a finite 
chain, and computed the critical ramping speed $c_{cr}$, which separates selftrapped localized from selftrapped delocalized 
states at the end of the ramping process. We find that $c_{cr}$ grows with increasing strength of disorder when the localization 
length is larger than the system size. That increase is due to the fact, that disorder is removing selection rules for the 
interaction of normal modes, present in the case of zero disorder. For strong disorder, i.e. when the localization length becomes 
smaller than the system size, the critical velocity drops with further increase of the disorder strength. That is due to the 
fact, that the normal modes are less and less spread, interact with
fewer other modes, and the nonlinearity induced frequency shift is more efficient in tuning the wavepacket faster into a selftrapped state.

\section{Acknowledgements}

We thank B. Altshuler, S. Komineas, D. Krimer and Ch. Skokos for useful discussions.

\end{document}